%% file: main.tex
\documentclass[10pt,conference]{IEEEtran}
\IEEEoverridecommandlockouts
\usepackage{url}

\ifCLASSINFOpdf
   \usepackage[pdftex]{graphicx}
\else
\fi

\usepackage{color}
\usepackage{ifthen}
\usepackage{soulutf8} 
\usepackage {bsymb,b2latex}
\usepackage{framed}
\usepackage{Event-B}
\usepackage{multirow}
\usepackage{longtable}
\usepackage{graphicx}
\usepackage{tabularx}
\usepackage{float}
\usepackage{flushend} 
\usepackage{cite}
\usepackage{amsmath,amssymb,amsfonts}
\usepackage{algorithmic}
\usepackage{graphicx}
\usepackage{textcomp}
\usepackage{xcolor}
\def\BibTeX{{\rm B\kern-.05em{\sc i\kern-.025em b}\kern-.08em
    T\kern-.1667em\lower.7ex\hbox{E}\kern-.125emX}}

\newboolean{showcomments}
\setboolean{showcomments}{true} 
\ifthenelse{\boolean{showcomments}}
{ \newcommand{\mynote}[2]{
      \fbox{\bfseries\sffamily\scriptsize#1}
        {\small\#{{{#2}\bf }}\#}}}
        { \newcommand{\mynote}[2]{}}

\begin{document}
%

\title{From \textit{What} to \textit{How}:\\ 
A Taxonomy of Formalized Security Properties}


\author{\IEEEauthorblockN{Imen Sayar}
\IEEEauthorblockA{
\textit{University of Toulouse/IRIT}\\
Toulouse, France \\
imen.sayar@irit.fr}
\and
\IEEEauthorblockN{Nan Messe}
\IEEEauthorblockA{
\textit{University of Toulouse/IRIT}\\
Toulouse, France \\
nan.messe@irit.fr}
\and
\IEEEauthorblockN{Sophie Ebersold}
\IEEEauthorblockA{
\textit{University of Toulouse/IRIT}\\
Toulouse, France \\
sophie.ebersold@irit.fr}
\and
\IEEEauthorblockN{Jean-Michel Bruel}
\IEEEauthorblockA{
\textit{University of Toulouse/IRIT}\\
Toulouse, France \\
jean-michel.bruel@irit.fr}}


%


\maketitle

\begin{abstract}
Confidentiality, integrity, availability, authenticity, authorization, and accountability are known as security properties that secure systems should preserve. They are usually considered as security final goals that are achieved by system development activities, either in a direct or an indirect manner. However, these security properties are mainly elicited in the high-level requirement phase during the System Development Life Cycle (SDLC) and are not refined throughout the latter phases as other artifacts such as attacks, defenses, and system assets. To align security properties refinement with attacks, defenses, and system assets refinements, we propose an SDLC taxonomy of security properties that may be used in a self-adaptive context and present the methodology for defining it. To verify and check the correctness of the resulting taxonomy, we use the Event-B formal language. 
\end{abstract}

\begin{IEEEkeywords}
taxonomy, formal methods, cybersecurity requirement, early verification and validation, security-critical system
\end{IEEEkeywords}


%
\IEEEpeerreviewmaketitle

\input{1introduction}

\input{2objective}

\input{3process}
\input{4taxonomy}

\input{5validation}
\input{6relatedwork}

\input{7discussion}
\input{8conclusion}






%

\bibliographystyle{plain}
\bibliography{0bib}




\end{document}

%% file: 1introduction.tex
\section{Introduction}\label{sec:into}
A system's security properties, e.g., confidentiality, integrity, and availability (the CIA triad), are to be ensured during different stages of a System Development Life Cycle (SDLC) to help system engineers build trustworthy systems. During the requirement phase, security properties are included in various activities, e.g., attack trees~\cite{widel2019beyond}, abuse cases, and misuse cases~\cite{hope2004misuse}, to help requirement analysts elicit security requirements. Then, during the design phase, different architectural measures, e.g., security patterns~\cite{schumacher2013security}, assurance cases~\cite{gacek2014resolute}, and security controls can be applied by architects as architecture decisions to satisfy security requirements, which in turn ensure security properties. Those architecture decisions are later implemented by developers during the implementation phase, where codes are also tested, using for example code review~\cite{braz2022less}, penetration testing~\cite{arkin2005software} and SAST~\cite{yang2019towards} to identify any security requirement's non-satisfaction or security property violation. Finally, systems in operation are monitored to detect security anomalies that go against security properties and are maintained or recovered in case of security breaches.

As we can see, security properties are addressed in various security activities throughout the SDLC, either directly or indirectly. However, the gains made by their inclusion in different stages are not equivalent. Dealing with security properties during the design phase and addressing them during the system operation entail different costs. Indeed, the earlier the security properties are covered and verified in system development, the earlier the vulnerabilities are detected. This results in less cost to be spent on incident responses and less the reputation of the company will be damaged. Therefore, security should not be an afterthought but be built into the system from the ground up. Identifying and verifying security properties since the early phases of SDLC, especially in the requirement and design phases, are thus critical for building a successful and secure system product.

\begin{figure}[!ht]
    \centering
    \includegraphics[scale=0.42]{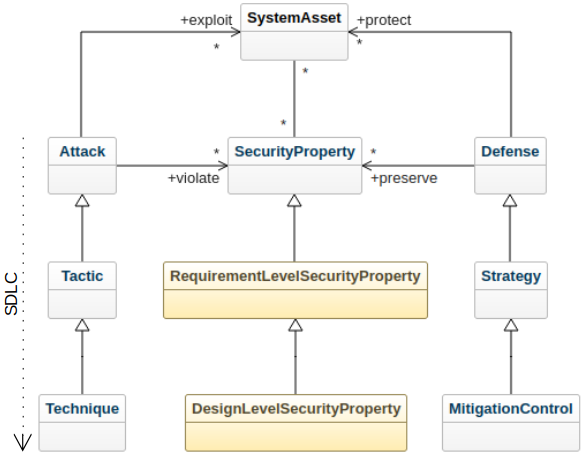}
    \caption{Security artifacts refinement model}
    \label{fig:model}
\end{figure}

In the information security practice and academic literature, security properties are known as \textbf{CIA} and \textbf{AAA}, which respectively refer to Confidentiality, Integrity, Availability and Authentication, Authorization, Accountability~\cite{iso2018iso, matulevivcius2017fundamentals, nweke2017using}. They serve as security final goals to be addressed in development activities~\cite{ma2008information, von2013information}. As illustrated in the white part of Figure~\ref{fig:model}, attacks, exploiting system assets and violating security properties, as well as defenses, protecting system assets and preserving security properties, can each be refined into attack tactics and defense strategies respectively, which are both addressed during the requirement and high-level design phases. These artifacts are then, in their turn, refined into attack techniques and mitigation controls correspondingly, which are dealt with during the lower-level design and implementation phases. This refinement is carried out throughout the SDLC. However, this is not the case for security properties due to their high-level nature and lack of support. Refining, tracing, and verifying the security properties during the different stages of the SDLC are therefore complicated tasks, which require dedicated support and techniques.

For instance, the security requirement ``\textit{The messages in transit between the client and the server should not be modified by a third party}." is elicited by targeting the integrity objective. Refining, representing, and satisfying this security requirement during the architecture design would result in several architecture decisions, such as protecting interfaces, supporting user-input validations, identifying race conditions, etc. In practice, determining those architecture decisions requires domain expertise, security expertise, and tool support, to capture all intended security properties that were considered during the requirement phase~\cite{verdon2004risk, yskout2020threat}. For example, in the widely-used threat modeling activity, domain experts and security experts need to work together in a brainstorming session to identify vulnerabilities and threats that may violate the security requirements, and provide mitigation controls which will be rendered as architecture decisions~\cite{messe2020asset}. Consequently, the resulting architectural decisions are highly dependent on human expertise and the chosen tools.

Moreover, in each phase of the SDLC, and particularly in the requirements definition phase, checking whether lower-level (e.g., component-level) requirements demonstrate the satisfaction of higher-level (e.g., system-level) requirements is still an area of ongoing research~\cite{whalen2012your}. Currently, even in safety-critical systems with well-understood domains, it is hard to specify requirements correctly~\cite{whalen2012your}.

To provide an SDLC support for aligning security properties refinement with attack and defense refinement, as shown in the yellow part of Figure~\ref{fig:model}, we propose a taxonomy of security properties in this paper. This taxonomy is described together with  the methodology used to define it.
The taxonomy presented in this paper addresses ``what'' security properties a system or a system element should preserve rather than ``how" to implement them. The proposed taxonomy can yet be bridged with lower-level implementation, which is not the focus of this paper.
To verify the proposed taxonomy, we use Event-B, a formal language, to early check its correctness. To the best of our knowledge, this is the first SDLC taxonomy of formalized security properties provided in the literature.  

This paper is organized as follows: Section~\ref{sec:objective} highlights the long-term objective of proposing the security properties taxonomy. Then, we elaborate in Section~\ref{sec:process} the tasks used for building and verifying this taxonomy with a BPMN-based process. In Section~\ref{sec:taxonomy}, the taxonomy of security properties is defined. To check its correctness, by  formally verifying and validating the taxonomy, we use Event-B. This is presented in Section~\ref{sec:validation}. Related works are discussed in Section~\ref{sec:relatedwork}. We then discuss the advantages of the taxonomy and identify its current limitation in Section~\ref{sec:discussion}. Finally, we conclude the paper and discuss future works in Section~\ref{sec:conclusion}.

%% file: 2objective.tex
\section{Taxonomy's Long-Term Objective}\label{sec:objective}

To illustrate the long-term objective of our security properties taxonomy, we use an analogy with the algorithm design paradigm. In computer science, a developer designs an algorithm and implements it into a program to solve a problem. When the problem becomes relatively complicated, the developer follows the ``divide and conquer" strategy~\cite{bentley1980multidimensional} to recursively break down the problem into two or more sub-problems of the same or related type, until these problems become simple enough to be solved directly.
With the aim of building secure systems, we pursue a similar philosophy to define a taxonomy of security properties. This taxonomy can help break down the high-level security objectives into sub-security objectives while aligning with SDLC phases. 

\begin{figure}
    \centering
    \includegraphics[scale=0.4]{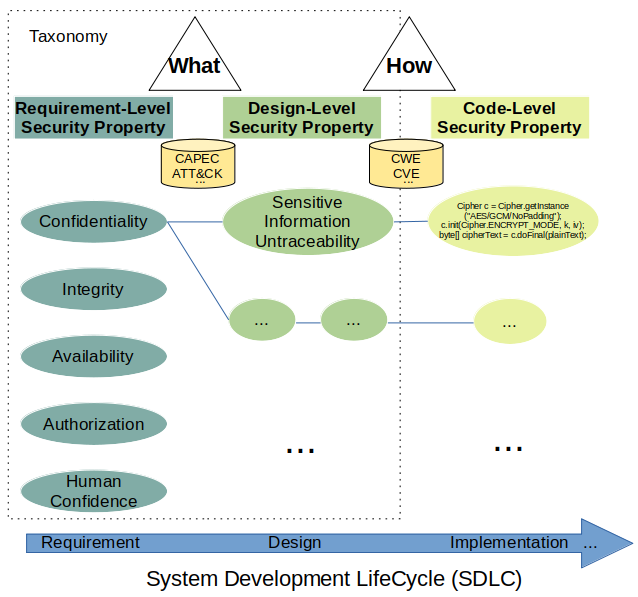}
    \caption{The long-term objective of the security properties taxonomy}
    \label{fig:bp}
\end{figure}

As shown in Figure~\ref{fig:bp}, the taxonomy begins with the reorganization of first-level security properties in the literature, e.g., ``confidentiality", which are final security goals that are usually addressed during the requirement phase. They tell thus ``what" the system should preserve. These requirement-level security properties are then decomposed and refined into second-level security properties to guide high-level design decisions, e.g., ``sensitive information untraceability". At this stage, it is still the ``what" that is addressed by architecture elements. This activity is iterated, hence there may be several levels of security properties. For example, ``sensitive information untraceability" can have the sub-properties ``networking protocol untraceability" and ``indicator untraceability". It is possible to continue this decomposition and refinement until low-level design activities, which allows transiting to ``how" to preserve those properties. Finally, those low-level design decisions are implemented at the code level, preserving thus code-level security properties. 

In this paper, we propose a taxonomy of security properties that identifies "what" the system should preserve, as shown in the left part of Figure~\ref{fig:bp}. 
Requirement analysts can use this taxonomy to elicit component-level security requirements, while architects can leverage the taxonomy to add security measures for architecture elements. In this way, when developers implement the code, those security measures can be smoothly transformed into code, while preserving the code-level security properties.

Security properties collected in this taxonomy are mainly extracted from CAPEC\footnote{https://capec.mitre.org/} and ATT\&CK\footnote{https://attack.mitre.org/}, and lately, CWE\footnote{https://cwe.mitre.org/}, CVE\footnote{https://cve.mitre.org/} will be leveraged for transiting into the ``how" part of Figure~\ref{fig:bp}. All of the above repositories are provided by MITRE\footnote{https://www.mitre.org/}. These repositories are widely used and frequently maintained by security practitioners and researchers. The advantages of CAPEC and ATT\&CK are their common and hierarchical features. They include common security attack patterns, which can be reused in various specific domains. Meanwhile, they organize the patterns into a hierarchy following different levels of abstraction, which eases the task of including those patterns in different SDLC phases.

%% file: 3process.tex
\section{Process of Extracting and Verifying Security Properties}~\label{sec:process} 

\begin{figure*}[!ht]
    \centering
    \includegraphics[width=\textwidth]{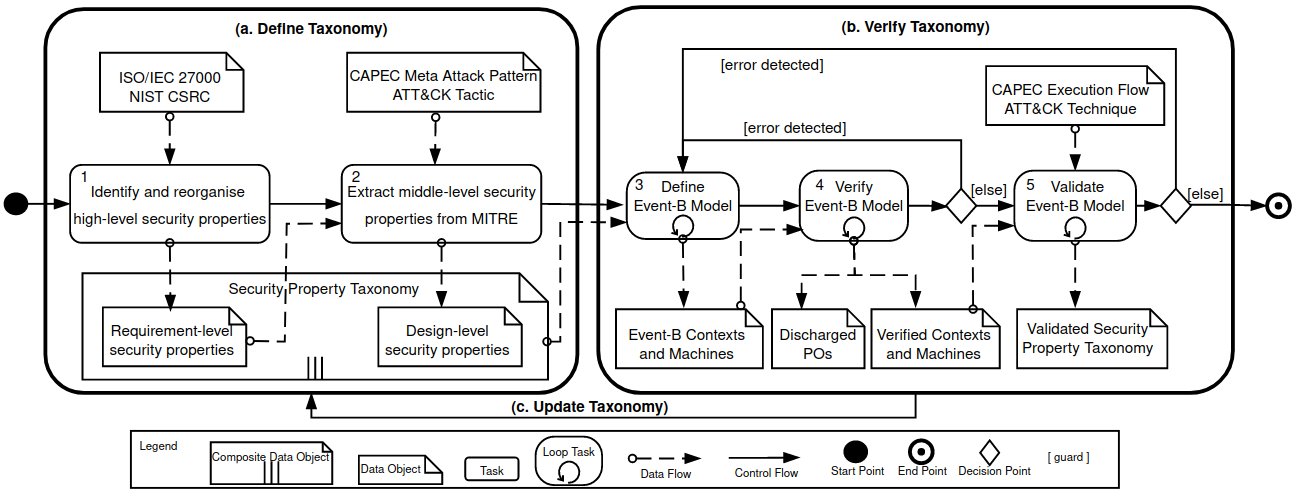}
    \caption{Extracting and verifying security properties process}
    \label{fig:process}
\end{figure*}

As mentioned in Section~\ref{sec:objective}, we rely on several MITRE repositories to build the taxonomy of security properties. In this section, we present a BPMN-based process to illustrate the taxonomy-building activities. In addition, the taxonomy needs to be verified before being reused by other practitioners and researchers. Therefore, the verifying activities are also included in this process. As shown in Figure~\ref{fig:process}, the extracting (part a) and verifying (part b) activities to build the taxonomy of security properties are described as follows:
\begin{enumerate}
    \item Firstly, high-level security properties are identified and reorganized from standards, such as ISO/IEC 27000 and NIST Computer Security Resource Center (CSRC)\footnote{https://csrc.nist.gov/}. This activity outputs requirement-level security properties, which constitute the top level of the taxonomy;
    \item Based on these top-level security properties, we extract their sub-security properties using CAPEC's Meta Attack Patterns\footnote{https://capec.mitre.org/data/definitions/282.html} and ATT\&CK Tactics\footnote{https://attack.mitre.org/tactics/}, both of which capture abstract characterization of a specific tactic or technique used in an attack. They are void of a specific technology or implementation and are meant to provide an understanding of a high-level approach. As a result, (several) design-level security properties are added as children of requirement-level security properties and the taxonomy is enriched;
    \item Once the taxonomy is defined, we  need to check its correctness.  To do that, we \textit{formalize} each property using the Event-B language~\cite{abrial10a}. We define a model composed of \texttt{MACHINEs} and \texttt{CONTEXTs} reflecting security properties. Our choice of this language is justified by its powerful constructions and the availability of tools used to define, verify, and validate models. Even though we have chosen Event-B, other languages like LTL~\cite{Pnueli77} or TLA+\cite{Lamport2002} can also be used to check the correctness of the taxonomy;   
    \item The defined Event-B models have to be \textit{verified}, \textit{i.e.,} built correctly. This activity is fully automated using the Rodin platform provers~\cite{abrial2010rodin}.  The outcome of the verification is a set of discharged Proof Obligations (POs) showing that the models are well-defined. If errors are detected, then we come back to the model and correct it;  
    \item Proving that the model is built correctly does not mean that it works correctly, \textit{i.e.,} it behaves as expected. This is the goal of the fifth step, during  which we \textit{validate} the formalized properties using animation and model-checking. This activity may detect errors and gives feedback, allowing the correction of the defined model.
    \end{enumerate}

 It is worth noting that this process is iterated. After defining, formalizing, verifying, and validating the security properties, our taxonomy may evolve by either adding new properties or modifying the existing taxonomy. In this case, we come back to \textit{update} the existing taxonomy. Such activity is described by the part \textit{c. Update Taxonomy} in Figure~\ref{fig:process}. 

%% file: 4taxonomy.tex
\section{Taxonomy of Security Properties}\label{sec:taxonomy}
\input{taxo}

In this section, we give the definition of each security property collected in the taxonomy, following the extracting process described in Section~\ref{sec:process}. For each security property, we point out the sources where it is extracted. The sources are shown by the \textit{name} of the repository and their \textit{ID}. For example, CAPEC-112 means the attack pattern 112 in CAPEC, and TA0043 means the tactic 0043 in ATT\&CK. The complete taxonomy showing the hierarchy of security properties can be found in Table~\ref{tab:taxo}.

\begin{itemize}
    \item \textbf{Confidentiality}: Assurance that information (in storage, in processing, and in transit) is not disclosed to unauthorized persons, processes, or devices. \textit{Sources: TA0043, TA0010, TA0036}
    \begin{itemize}
        \item \textbf{Sensitive information untraceability}: Sensitive information (e.g., passwords, encryption keys, database lookup keys, initial values to one-way functions, etc.) should not be discovered or gathered by an attacker. \textit{Sources: CAPEC-112, TA0009, TA0035, TA0100}
        \begin{itemize}
            \item \textbf{Networking protocol untraceability}: The function and characteristics of a communication protocol, e.g., data transmission syntax and meaningful content, including packet or content delimiters used by the protocol, should not be revealed by an attacker. \textit{Sources: CAPEC-192}
            \item \textbf{Indicator untraceability}: Indicators that uniquely identify specific details about the target, such as operating system and application versions, should not be discovered or gathered by an attacker. \textit{Sources: CAPEC-224, TA0007, TA0032, TA0102}
        \end{itemize}
        \item \textbf{Opacity}: The discovered secret (sensitive information) of the system should not be diffused by an intruder (attacker)  who is not the intended recipient. \textit{Sources: CAPEC-560}
        \item \textbf{System property untraceability}: An attacker should not infer the system's properties such as its actual behavior, its security mechanisms, configuration, or potential vulnerabilities based on the information flow, which includes open ports, applications, and their versions, network topology, and similar information. \textit{Sources: CAPEC-28, CAPEC-116, CAPEC-117, CAPEC-169, CAPEC-188}
    \end{itemize}
    \item \textbf{Integrity}: System assets (data/functionality) have not been altered or misused, respectively, in an unauthorized manner. \textit{Sources: CAPEC-184, CAPEC-212, TA0040, TA0034, TA0105}
    \begin{itemize}
        \item \textbf{Non-repudiation}: Assurance that the sender of information is provided with proof of delivery and the recipient is provided with proof of the sender’s identity, so neither can later deny having processed the information.
        \item \textbf{Authenticity}: Data originating from its purported source; or any resource that is associated with an identity (human or non-human) should not be stolen or spoofed. \textit{Sources: CAPEC-94, CAPEC-151}
        \begin{itemize}
            \item \textbf{Identifier trustfulness}: A trusted identifier (e.g., session ID, resource ID, cookie, etc.) should not be guessed, obtained, or “ridden” by an attacker. \textit{Sources: CAPEC-21, CAPEC-240, CAPEC-161, TA0006, TA0031}
            \item \textbf{Client trustfulness}: The client with which the server communicates should be a valid client. \textit{Sources: CAPEC-22, TA0011, TA0037, TA0101}
            \item \textbf{Action trustfulness}: One action should not be disguised for another by an attacker, which tricks a user into initiating one type of action when they intend to initiate a different action. \textit{Sources: CAPEC-173}
        \end{itemize}
        \item \textbf{Network infrastructure integrity}: A network infrastructure characteristics, such as the routing of network messages, the communication protocol, or a setting or parameter on a communication channel, should not be manipulated by an attacker. Consequence: information gathering on network objects or effects a change in the ordinary information flow between network objects. \textit{Sources: CAPEC-216, CAPEC-272, CAPEC-594, TA0001, TA0027, TA0038, TA0108}
        \item \textbf{Zero race condition}: An attacker should not be able to leverage a race condition by “running the race”, modifying the resource, and modifying the normal execution flow. \textit{Sources: CAPEC-26}
        \item \textbf{User state information integrity}: User state information (e.g., username, payment information, browsing history, and application-specific contents such as items in a shopping cart) maintained by the target software should not be modified by an attacker. \textit{Sources: CAPEC-74}
        \item \textbf{Interface integrity}: The interface (e.g., Application Programming Interface (API) or System-on-Chip (SoC)) functionality of the target system should not be discovered and manipulated by an attacker. \textit{Sources: CAPEC-113}
        \item \textbf{Buffer space integrity}: Data in a buffer should not be retrieved or overwritten by any unintended programmatic means. \textit{Sources: CAPEC-123}
        \item \textbf{Shared resource integrity}: A resource shared between multiple applications, an application pool or hardware pin multiplexing, should not be manipulated by an attacker. \textit{Sources: CAPEC-124}
        \item \textbf{Pointer integrity}: Pointer variables should not be manipulated by an attacker to access an unintended memory location. \textit{Sources: CAPEC-129}
        \item \textbf{Content integrity}: The content (e.g., web pages, email messages, file transfers, or the content of other network communication protocols) should not be modified by an attacker. \textit{Sources: CAPEC-148}
        \item \textbf{Input processing correctness}: The supplying of the user input should respect standard and expected form, like that, the attacker can not control the format, structure, and data composition of inputs. \textit{Sources: CAPEC-153, CAPEC-242, CAPEC-248, CAPEC-586}
        \begin{itemize}
            \item \textbf{Parameter integrity}: The content of a request parameter should not be manipulated by an attacker. \textit{Sources: CAPEC-137}
        \end{itemize}
        \item \textbf{Resource trustfulness}: The characteristics of a resource should not be modified.
        \begin{itemize}
            \item \textbf{Resource location trustfulness}: An attacker should not be able to deceive an application or user and convince them to request a resource from an unintended location. \textit{Sources: CAPEC-154, CAPEC-175}
            \item \textbf{Resource metadata integrity}: The metadata of a resource (e.g., file, directory, repository, etc.) should not be altered by an attacker. \textit{Sources: CAPEC-690}
        \end{itemize}
        \item \textbf{File integrity}: File contents or attributes, such as extensions or names, should not be modified by an attacker, which could cause incorrect processing by an application. \textit{Sources: CAPEC-165}
        \item \textbf{External configuration and environment trustfulness}: External configuration files and libraries should not be modified by an attacker to affect the system’s behavior. \textit{Sources: CAPEC-176}
        \item \textbf{Supply chain integrity}: A technology, product, or component should not be modified in the supply chain.
        \begin{itemize}
            \item \textbf{Manufacture integrity}: A technology, product, or component should not be modified during its manufacturing stage. \textit{Sources: CAPEC-438}
            \item \textbf{Distribution integrity}: A technology, product, or component should not be modified during its distribution stage. \textit{Sources: CAPEC-439}
            \item \textbf{Deployment integrity}: A technology, product, component, a subcomponent, or a new one installed should not be modified during its deployment stage. \textit{Sources: CAPEC-440, TA0039}
            \item \textbf{Operation integrity}: Malicious code, such as malware, rootkits, ransomware, spyware, and adware, should not be installed or executed during the product’s operation stage. \textit{Sources: CAPEC-441, CAPEC-549, TA0002, TA0041, TA0104}
        \end{itemize}
        \item \textbf{Hardware non-interference}: Electronic device should not be interfered by disruptive signals or events, or alters the physical environment the device operates in. \textit{Sources CAPEC-624}
    \end{itemize}
    \item \textbf{Availability}: the property that information is accessible and usable upon demand by an authorized person. \textit{Sources: CAPEC-548, CAPEC-607, TA0040, TA0034, TA0107}
    \begin{itemize}
        \item \textbf{Deadlock freeness}: A target software stocking, transiting, or using information should not have a deadlock condition. Examples of a deadlock condition are: Mutual exclusion; hold and wait or resource holding; no pre-emption; circular wait~\footnote{https://en.wikipedia.org/wiki/Deadlock}. \textit{Sources: CAPEC-25}
        \item \textbf{Resource controllability}: The number of requests to a system’s resource in a given period of time should be controlled so that legitimate users can always access the service. \textit{Sources: CAPEC-125}
        \item \textbf{Resource allocation moderation}: Allocation of finite resources, such as memory, bandwidth, processing cycles, etc., should be moderate. \textit{Sources: CAPEC-130}
        \item \textbf{Zero resource leak}: An attacker should not be able to utilize a resource leak on the system to delete the quantity of the resources available to service legitimate requests. \textit{Sources: CAPEC-131}
        \item \textbf{Engaging time controllability}: The time for a resource engagement upon a request should be controllable to avoid sustained client engagement. \textit{Sources: CAPEC-227}
    \end{itemize}
    \item \textbf{Authorization}: Right or permission that is granted to a system entity to access a system resource. \textit{Sources: TA0003, TA0028, TA0110}
    \begin{itemize}
        \item \textbf{Authentication mechanism faultlessness}: An authentication mechanism should not have or expose inherent weakness in its implementation. \textit{Sources: CAPEC-114}
        \item \textbf{Authentication unavoidability}: An authentication mechanism should not be able to be evaded or circumvented. \textit{Sources: CAPEC-115}
        \begin{itemize}
            \item \textbf{Physical authentication unavoidability}: The building security and the surveillance should be ensured with secure entry points, and the electronic or physical locks should not be bypassed. \textit{Sources: CAPEC-390, TA0106}
        \end{itemize}
        \item \textbf{Privilege preservation}: A system’s features should be reserved for privileged users or administrators and are not used by lower or non-privileged accounts because of the access control mechanisms’ absence or misconfiguration. \textit{Sources: CAPEC-122, CAPEC-233, CAPEC-554, TA0004, TA0029, TA0111}
        \item \textbf{Physical protection}: A physical device should not be stolen by an attacker to gain physical access to a system. \textit{Sources: CAPEC-507}
    \end{itemize}
    \item \textbf{Human confidence}: The legitimate individual should not divulge useful information to an untrusted attacker or perform actions that serve the attacker’s interests. \textit{Sources: CAPEC-410, CAPEC-416}
\end{itemize}

%% file: taxo.tex

\begin{table*}[h]
\centering
\resizebox{\textwidth}{!}{
\begin{tabular}{|lll|}
\hline
\multicolumn{3}{|c|}{\textbf{What}} \\ \hline
\multicolumn{1}{|c|}{\textbf{\begin{tabular}[c]{@{}c@{}}First-Level\\ Security Property\end{tabular}}} &
  \multicolumn{1}{c|}{\textbf{\begin{tabular}[c]{@{}c@{}}Second-Level\\ Security Property\end{tabular}}} &
  \multicolumn{1}{c|}{\textbf{\begin{tabular}[c]{@{}c@{}}Third-Level\\ Security Property\end{tabular}}} \\ \hline
\multicolumn{1}{|l|}{\multirow{4}{*}{1. Confidentiality}} &
  \multicolumn{1}{l|}{\multirow{2}{*}{1.1. Sensitive Information Untraceability}} &
  1.1.1. Networking Protocol Untraceability \\ \cline{3-3} 
\multicolumn{1}{|l|}{} &
  \multicolumn{1}{l|}{} &
  1.1.2. Indicator Untraceability \\ \cline{2-3} 
\multicolumn{1}{|l|}{} &
  \multicolumn{2}{l|}{1.2. Opacity} \\ \cline{2-3} 
\multicolumn{1}{|l|}{} &
  \multicolumn{2}{l|}{1.3. System Property Untraceability} \\ \hline
\multicolumn{1}{|l|}{\multirow{22}{*}{2. Integrity}} &
  \multicolumn{2}{l|}{2.1. Non-repudiation} \\ \cline{2-3} 
\multicolumn{1}{|l|}{} &
  \multicolumn{1}{l|}{\multirow{3}{*}{2.2. Authenticity}} &
  2.2.1. Identifier Trustfulness \\ \cline{3-3} 
\multicolumn{1}{|l|}{} &
  \multicolumn{1}{l|}{} &
  2.2.2. Client Trustfulness \\ \cline{3-3} 
\multicolumn{1}{|l|}{} &
  \multicolumn{1}{l|}{} &
  2.2.3. Action Trustfulness \\ \cline{2-3} 
\multicolumn{1}{|l|}{} &
  \multicolumn{2}{l|}{2.3. Network Infrastructure Integrity} \\ \cline{2-3} 
\multicolumn{1}{|l|}{} &
  \multicolumn{2}{l|}{2.4. Zero Race Condition} \\ \cline{2-3} 
\multicolumn{1}{|l|}{} &
  \multicolumn{2}{l|}{2.5. User State Information Integrity} \\ \cline{2-3} 
\multicolumn{1}{|l|}{} &
  \multicolumn{2}{l|}{2.6. Interface Integrity} \\ \cline{2-3} 
\multicolumn{1}{|l|}{} &
  \multicolumn{2}{l|}{2.7. Buffer Space Integrity} \\ \cline{2-3} 
\multicolumn{1}{|l|}{} &
  \multicolumn{2}{l|}{2.8. Shared Resource Integrity} \\ \cline{2-3} 
\multicolumn{1}{|l|}{} &
  \multicolumn{2}{l|}{2.9. Pointer Integrity} \\ \cline{2-3} 
\multicolumn{1}{|l|}{} &
  \multicolumn{2}{l|}{2.10. Content Integrity} \\ \cline{2-3} 
\multicolumn{1}{|l|}{} &
  \multicolumn{1}{l|}{2.11. Input Processing Correctness} &
  2.11.1. Parameter Integrity \\ \cline{2-3} 
\multicolumn{1}{|l|}{} &
  \multicolumn{1}{l|}{\multirow{2}{*}{2.12. Resource Trustfulness}} &
  2.12.1. Resource Location Trustfulness \\ \cline{3-3} 
\multicolumn{1}{|l|}{} &
  \multicolumn{1}{l|}{} &
  2.12.2. Resource Metadata Integrity \\ \cline{2-3} 
\multicolumn{1}{|l|}{} &
  \multicolumn{2}{l|}{2.13. File Integrity} \\ \cline{2-3} 
\multicolumn{1}{|l|}{} &
  \multicolumn{2}{l|}{2.14. External Configuration and Environment Trustfulness} \\ \cline{2-3} 
\multicolumn{1}{|l|}{} &
  \multicolumn{1}{l|}{\multirow{4}{*}{2.15. Supply Chain Integrity}} &
  2.15.1. Manufacture Integrity \\ \cline{3-3} 
\multicolumn{1}{|l|}{} &
  \multicolumn{1}{l|}{} &
  2.15.2. Distribution Integrity \\ \cline{3-3} 
\multicolumn{1}{|l|}{} &
  \multicolumn{1}{l|}{} &
  2.15.3. Deployment Integrity \\ \cline{3-3} 
\multicolumn{1}{|l|}{} &
  \multicolumn{1}{l|}{} &
  2.15.4. Operation Integrity \\ \cline{2-3} 
\multicolumn{1}{|l|}{} &
  \multicolumn{2}{l|}{2.16. Hardware Non-interference} \\ \hline
\multicolumn{1}{|l|}{\multirow{5}{*}{3. Availability}} &
  \multicolumn{2}{l|}{3.1. Deadlock Freeness} \\ \cline{2-3} 
\multicolumn{1}{|l|}{} &
  \multicolumn{2}{l|}{3.2. Resource Controllability} \\ \cline{2-3} 
\multicolumn{1}{|l|}{} &
  \multicolumn{2}{l|}{3.3. Resource Allocation Moderation} \\ \cline{2-3} 
\multicolumn{1}{|l|}{} &
  \multicolumn{2}{l|}{3.4. Zero Source Leak} \\ \cline{2-3} 
\multicolumn{1}{|l|}{} &
  \multicolumn{2}{l|}{3.5. Engaging Time Controllability} \\ \hline
\multicolumn{1}{|l|}{\multirow{4}{*}{4. Authorization}} &
  \multicolumn{2}{l|}{4.1. Authentication mechanism faultlessness} \\ \cline{2-3} 
\multicolumn{1}{|l|}{} &
  \multicolumn{1}{l|}{4.2. Authentication Unavoidability} &
  4.2.1. Physical Authentication Unavoidability \\ \cline{2-3} 
\multicolumn{1}{|l|}{} &
  \multicolumn{2}{l|}{4.3. Privilege Preservation} \\ \cline{2-3} 
\multicolumn{1}{|l|}{} &
  \multicolumn{2}{l|}{4.4. Physical Protection} \\ \hline
\multicolumn{3}{|l|}{5. Human Confidence} \\ \hline
\end{tabular}
}
\caption{The taxonomy of security properties}
\label{tab:taxo}
\end{table*}

%% file: 5validation.tex
\section{Evaluation}
\label{sec:validation}
To evaluate the approach, we formalize the taxonomy using Event-B. In this section, we first describe the background needed to perform this evaluation process. Then, we show the application of this evaluation  using an example of the two properties of Integrity and Authenticity which we formalize, verify, and validate using Event-B. 

\subsection{Background}

\subsubsection{Formal Methods}
Formalizing properties is relevant when it comes to building correct systems.
The security requirements may be expressed and checked early in the development life cycle using formal methods like B~\cite{abrial05} and Event-B~\cite{abrial10a} languages. 

One interesting direction of this work is the expression of security properties using formal methods and V\&V tools available in the Rodin platform\footnote{\url{http://www.event-b.org/}}. Event-B is a formal method that supports the gradual construction of correct formal specifications through a proof activity. This method is an extension of the classical B language. It is used for the development of formal models for reactive systems and sequential and distributed algorithms. It also supports the development of so-called hybrid systems that combine software, hardware, and user interfaces.

\subsubsection{Refinement}
To manage the complexity when developing complex systems, we use the refinement technique. This concept was introduced in the early 1970s~\cite{Dijkstra76}. It is the technique of transforming an abstract model into a more detailed and concrete model. The latter must preserve the properties of the abstract model as well as its behavior. We can talk about refinement in different levels  of software or system development, notably upstream through the specification and downstream during implementation or coding. Wirth~\cite{Wirth71} used refinement at the programming level. In his approach, the refinement is considered as a sequence of design decisions that consists in decomposing the program's tasks into sub-tasks and the data into data structures. The refinement of formal specifications supports the incremental elaboration of a complex specification starting from a coherent abstract model and concretizing it step-by-step~\cite{sayar2020formalization}. This consistency is ensured via refinement proof obligations when using formal methods like Event-B. 

\subsubsection{Verification and Validation (V\&V)}
\paragraph{\textit{About the verification}}
In the Event-B paradigm, each component under construction must be mathematically correct. This is achieved through proof obligations denoted \textit{POs}. A \textit{PO} is  a predicate that defines a property that must be proven to ensure the internal consistency of a formal model and guarantee correct refinement. The POs are automatically generated by the proof obligation generators. They are then proved by provers such as the SMT~\cite{DeharbeFGV12} and SAT~\cite{EenS03} solvers. These provers are composed of a set of reasoners written in Java. Each reasoner provides a proof rule. The latter corresponds to a proof tree generated by a tool called the \textit{sequent calculator}. A proof tree is associated with each PO showing the path used to prove it.

\paragraph{\textit{About the validation}}
It aims to provide an answer to the question ``Are we developing the right software or system?" In other words, the goal is to make sure that the software or system developed meets the customer's needs. This activity, 
used throughout the formal development process, makes it possible to detect a set of problems such as deadlocks or unauthorized behaviors. There are various validation techniques such as :
\begin{itemize}
    \item \textit{Animation}, which facilitates exploring the different states that a model can have,
     \item \textit{model translation}, which consists in transforming models into executable programs,
    \item \textit{simulation}, through which a model can be executed and
    \item \textit{model-checking}, the idea which emerged in 1980~\cite{EmersonC80}. Its objective is to verify whether a given model satisfies a property. If not, the model-checker provides a counter-example.
\end{itemize}
In this work, we use the ProB animator model-checker tool~\cite{LeuschelProB}, which is able to design and simulate animation scenarios. This tool contains a powerful module called ProB Disprover \footnote{\url{https://prob.hhu.de/w/index.php?title=Tutorial_Disprover}} which gives counter-examples for failed POs.

\subsection{Taxonomy Formalization}
As mentioned earlier, we have used the Event-B formal method~\cite{abrial10a} to define, verify and validate the models. Two components characterize an Event-B model: the Context denoted by the keyword \texttt{CONTEXT} and the Machine denoted by the keyword \texttt{MACHINE}. While the context describes the static part of a model through constants and sets, the machine describes the dynamic one reflecting a state-transition model in which a state is described by the value of the variables, and the transitions are modeled through \texttt{Events}. A machine can refine, via the \texttt{REFINES} relation, only one other machine. It can use, via the \texttt{SEES} relation, explicitly or implicitly one or more contexts. A context can extend, through the \texttt{EXTENDS} keyword, one or more contexts. Figure~\ref{Integrity-model} gives an excerpt of the Event-B machine and context translating the \textit{Integrity} property. The state of the machine \texttt{Integrity\_Mch} is defined by its variables such as \texttt{asset}, \texttt{ask\_access}, and \texttt{assetStatus}. The types of these variables may be: (i) predefined ones like the \texttt{BOOL} type (for booleans) or (ii) new user types defined in the context \texttt{Integrity\_Ctx0} as the \texttt{SYSTASSET} referring to the set of system assets or \texttt{ASSETSATUS} describing the status of each asset (\texttt{same}, \texttt{modified} or \texttt{misused}). The transitions between different states of the machine \texttt{Integrity\_Mch} are defined through the \textit{events}. For instance, the event \texttt{ask\_modify\_asset} represents a request done by a user ---who may be an attacker--- to modify the state of a system asset. This request may be accepted or refused  (via the \texttt{refuse\_modify\_asset}) by the administrator of the system in question.  It is worth noting that in this formalized machine, we represent different kinds of events: (i) events  performed by a system user/attacker (e.g., \texttt{ask\_modify\_asset}), (ii)  those fulfilled by the system itself (e.g., \texttt{reinitialize}), and (iii) events done by the system administrator (e.g.,  \texttt{refuse\_modify\_asset})).    

\begin{figure}[!h]
\begin{center}
\fbox{
\begin{minipage}{0.45\textwidth}
 \input{Integrity_Ctx0.tex} 
\end{minipage}}
\vspace{0.2cm}
\fbox{
\begin{minipage}{0.45\textwidth}
 \input{integrity.tex} 
\end{minipage}}
\end{center}
\caption{Machine seeing a Context}
\label{Integrity-model}
\end{figure}

\begin{figure*}[!ht]
\centering
\includegraphics[width=5.8in]{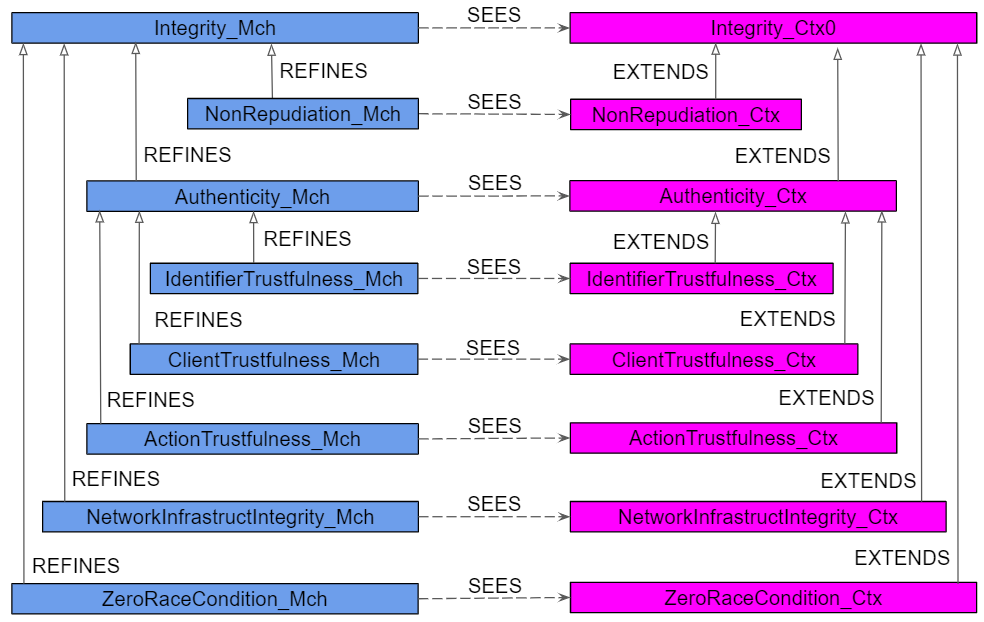}
\caption{The formalized taxonomy classification for the Integrity property and its descendants.}
\label{fig_integrity_arch}
\end{figure*}

The taxonomy supports linking \textit{what to protect} with \textit{how to protect}. The proposed taxonomy of security properties is initially extracted and defined using deep analysis of the CAPEC and the ATT\&CK repositories.  
To prove that this classification is correct, we formalized each classified property using the Event-B formal language. An overview of the produced Event-B models concerning integrity is given in Figure~\ref{fig_integrity_arch}. This viewpoint diagram is obtained using the Event-B Project Diagram\footnote{\url{https://wiki.event-b.org/index.php/Project_Diagram}} plugin. The machines are represented by the left (blue) boxes, which see and use the contexts represented by the right (pink) boxes. A powerful concept of the Event-B model is the refinement strategy. This concept ensures the correction of the proposed  taxonomy. In fact,  when a concrete machine like  \texttt{NonRepudiation\_Mch} refines another one such as \texttt{Integrity\_Mch}, this reflects that the \texttt{NonRepudiation\_Mch} gets all the properties of  \texttt{Integrity\_Mch} and does not contradict it. 

\subsection{Checking the Correctness of the Taxonomy}
Once defined, each model of Figure~\ref{fig_integrity_arch} is verified via the automatic provers of the Rodin platform and validated using the ProB animator and model-checker.    
Validation can only be effective if the requirements are precise. Precision in science and technology is typically achieved by using mathematical methods and notations, also known in software engineering as formal methods and notations~\cite{bruel2021role}. 

All the models are proven correct by construction. This means that the taxonomy is promising. For each model ---composed of a context and a machine--- the verification activity consists of generating Proof obligations (POs). Table~\ref{POs-for-inegrity} gives a summary of the generated POs for some of the formalized properties. In this table, we show that not all the POs were automatically discharged; in some cases, we interacted with the Rodin provers to help them to discharge the non-discharged POs. 

\begin{table}[!ht]
\begin{center}
\begin{tabular}{|p{0.24\textwidth}|p{0.055\textwidth}|p{0.06\textwidth}|p{0.025\textwidth}|}
\hline 
\textit{\textbf{Property model}} & \textit{\textbf{Automatic POs}} & 
 \textit{\textbf{Interactive POs}} & \textit{\textbf{Total}}\\
\hline 
\texttt{Integrity\_Mch} & 10 & 2 & 12  \\
\hline
\texttt{Authenticity\_Mch} & 11 & 2 & 13 \\
\hline
\texttt{Authenticity\_Ctx} & 1 & 1 & 2 \\
\hline
\texttt{IdentifierTrustfulness\_Mch} & 3 & 4 & 7 \\
\hline
\end{tabular}
\caption{Generated POs for some formalized properties}
\label{POs-for-inegrity}
\end{center}
\end{table}  

\begin{figure*}[!h]
\centering
\includegraphics[width=6.9in]{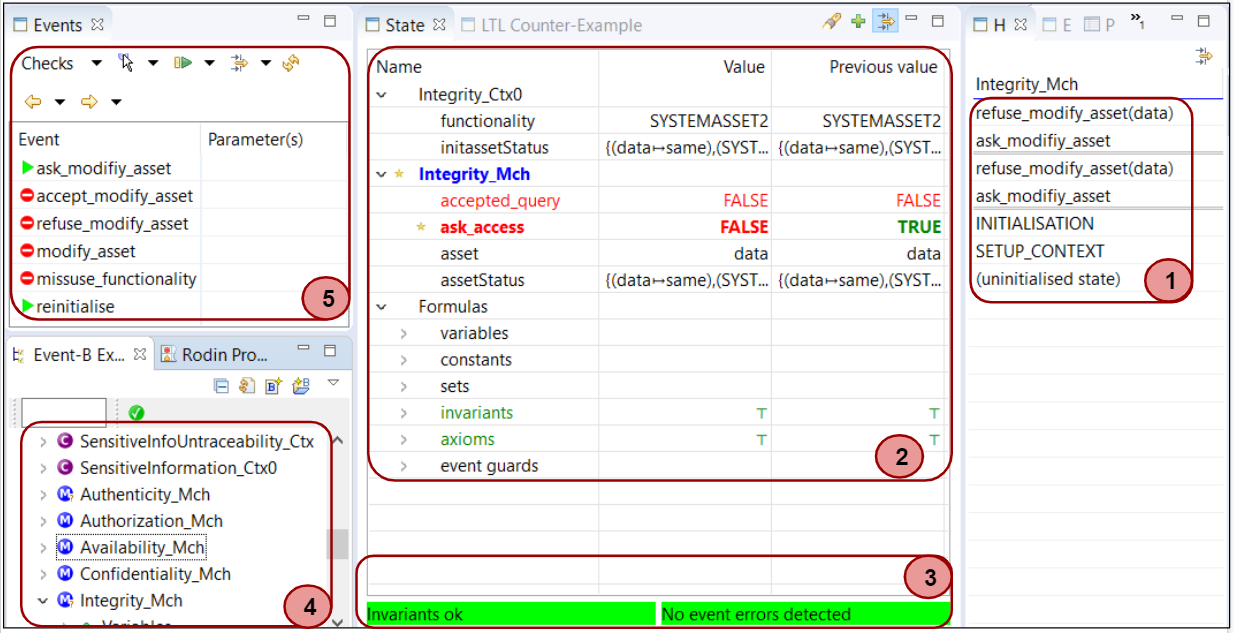}
\caption{Animation and model-checking of the Integrity property using the ProB tool.}
\label{fig_integrity_simulation}
\end{figure*}

The verification and validation methodology is based on three main steps: (i) formalize each security property into an Event-B model composed of a \texttt{Machine} seeing a \texttt{Context}, (ii) verify the formal model using the Rodin provers, and  (iii) validate this model using the ProB animator and model-checker. Figure~\ref{fig_integrity_simulation} shows an attack scenario for the Integrity property, in which an attacker tries to modify the assets of a system. In this figure, the following numbers have the following meanings: 
\begin{itemize}
    \item \textbf{1} refers to the actual state of the attack scenario execution. In the current case, the attacker tries firstly to ask for modifying assets (through the \texttt{ask\_modify\_asset} event). His/her request is refused (via the \texttt{refuse\_modify\_asset} event). The attacker retried again, and the formal model makes this attempt impossible. This proves that the model is secure against this attack scenario.
    \item \textbf{2} shows the state of the current \texttt{Integrity\_Mch} machine and its \texttt{Integrity\_Ctx0} context. This state is defined through the values associated with each variable, constant, invariant, axiom, and event guards. 
    \item \textbf{3} indicates when the model invariants are violated or events have errors.
    \item \textbf{4} mentions the proposed Event-B project structure in terms of machines and contexts.
    \item \textbf{5} shows the current machine events  that may be triggered (preceded  by the green triangle symbol) and those that are not accessible (preceded by the red circle symbol).   
\end{itemize}

\begin{figure}[!h]
\centering
\includegraphics[width=3.5in]{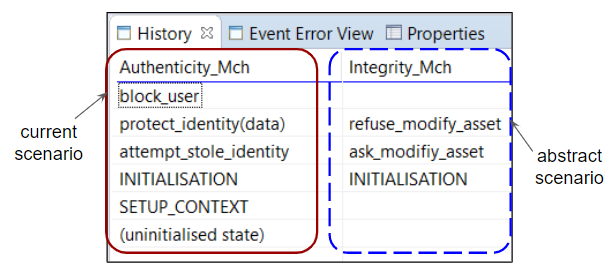}
\caption{Concrete (left side of the figure) and abstract (right side) simulation scenarios for the Authenticity and the Integrity properties, respectively.}
\label{fig_integrity_authent_animation}
\end{figure}

While animating and model-checking a concrete model using a scenario, an abstract scenario is automatically checked on the abstract machine. For example, Figure~\ref{fig_integrity_authent_animation} shows the animation of the machine \texttt{Authenticity\_Mch} using a scenario on the left side of the figure. At the same time, since this machine is refining the \texttt{Integrity\_Mch}, the sequence of the abstract events \texttt{INITIALISATION} \texttt{ask\_modify\_asset} and \texttt{refuse\_modify\_asset} is animated. This is relevant to show that the behavior of the concrete machine is respecting the behavior of the abstract one. This concludes that this step of the taxonomy definition (Authenticity is a sub-property of the Integrity property) is correct.   

%% file: Integrity_Ctx0.tex
\textbf{CONTEXT} \textit{Integrity\_Ctx0}
\vfill  \textbf{SETS}
    \vfill \hspace{0.2cm} \textit{SYSTASSET}, \textit{ASSETSSTATUS}
\vfill  \textbf{CONSTANTS}
	\vfill \hspace{0.2cm} \textit{data}, \textit{functionality}, \textit{same}, \textit{modified}, \textit{misused}, \textit{\ldots}
\vfill \textbf{AXIOMS} 
	\vfill  \hspace{0.2cm}  \textit{axm1: data $\in$ SYSTASSET}
	\vfill \hspace{0.2cm}  \textit{axm2: functionality $\in$ SYSTASSET} 
        \vfill \hspace{0.2cm}  \textit{axm3: partition(ASSETSSTATUS, {same}, {modified}, {misused})}	
\vfill \textbf{END}

%% file: integrity.tex
\textbf{MACHINE} \textit{Integrity\_Mch}
\vfill  \textbf{SEES}  \textit{Integrity\_Ctx0}
\vfill  \textbf{VARIABLES}
	\vfill \hspace{0.2cm} \textit{asset}, \textit{ask\_access}, \textit{assetStatus}, \textit{\ldots}
\vfill \textbf{INVARIANTS} 
	\vfill  \hspace{0.2cm}  \textit{inv1: asset $\in$ SYSTASSET}
	\vfill \hspace{0.2cm}  \textit{inv2: ask\_access $\in$ BOOL} 
        \vfill \hspace{0.2cm}  \textit{inv3: assetStatus $\in$ SYSTASSET $\leftrightarrow$ ASSETSSTATUS}	
\vfill  \textbf{EVENTS}
	\vfill \hspace{0.2cm} \textbf{INITIALISATION}
	\vfill \hspace{0.2cm} \textit{ask\_modify\_asset} // user/attacker event    
  	\vfill \hspace{0.2cm} \textit{refuse\_modify\_asset} // administrator event
        \vfill \hspace{0.2cm} \textit{reinitialise} // system event
        \vfill \hspace{0.2cm} \textit{\ldots}
\vfill \textbf{END}

%% file: 6relatedwork.tex
\section{Related Work}\label{sec:relatedwork}
\label{sec:related-work}

Some classifications of security properties have been presented in the literature~\cite{focardi2000classification, aldini2006classification, focardi1994taxonomy} and have been mathematically proven. However, these classifications, e.g., secrecy, non-interference, and authenticity, are equivalent to the first-level security properties in the taxonomy we propose here.
To the best of our knowledge, our paper introduces the first SDLC taxonomy of formalized security properties. 

The first-level security properties in this taxonomy usually serve as security objectives in practice. According to Souag. et al.~\cite{souag2016reusable}, no taxonomy of security objectives has been proposed in the literature (cf. Table 2~\cite{souag2016reusable}). 
However, those security properties have already been formalized. We thus present in this section existing works about the formalization of security properties.

The idea of formalizing the security properties is tackled by Stock et al. in~\cite{stock2022application} where they propose a validation obligation (VO) approach that is applied for security requirements. These requirements later are expressed in Event-B \texttt{machines} and \texttt{contexts} and concern specifically the authentication property. They write Validation Tasks (VT) and use views to help stakeholders understand, extract knowledge and have feedback from these views supporting the improvement of models and security properties. 

Rouland et al.~\cite{Rouland2019} suggest a (technology-independent) approach to specify security requirements, model them and verify them using a set of security policies. They use the  Alloy modeling language~\cite{JacksonAlloy} to guarantee the respect of the security properties. They are interested in the Confidentiality, Integrity, and Availability security properties in component-based software architecture models. 
Starting from informal security requirements, the authors formalize the security properties CIA. They model the system to check in Alloy and then formalize the security properties in the form of Alloy \texttt{Predicate/Assertion/Fact} constructions. Then, they check if the system respects the described properties using a command: \texttt{check seqReq for n} (meaning that they check if  the system respects the seqReq for n steps/iterations). If the output of the last step is OK, it is certain that the system respects the given safety property. Otherwise, the Alloy Analyzer gives a counterexample meaning the violation of such property. In this case, the developer defines a security policy to enforce the given security property. In our work, we choose to use Event-B because it is more user-friendly than Alloy and allows other practitioners or researchers to easily follow the approach we propose here. 

It is worth noting that none of the above works provides an SDLC and formalized taxonomy of security properties. They mainly focus on the formalization of the first-level security properties in our provided taxonomy.

%% file: 7discussion.tex
\section{Discussion}\label{sec:discussion}
\label{sec:discussion}
In this section, we discuss the advantages and the current limitations of the security properties taxonomy proposed in this paper.

Firstly, software engineers and security experts would be able to rely on this taxonomy for identifying threats and proposing defenses throughout the SDLC. In other words, this taxonomy helps to align security properties with attacks, defenses, and system assets, as shown in Figure~\ref{fig:model}. The sub-security properties can thus be mapped into sub-systems, components, and sub-components levels, thanks to this taxonomy. This taxonomy can further be included in a static model of interrelationship that ties security requirements with architectural elements, and such a mapping is equally essential for both building and verifying a system.

Secondly, our taxonomy is promising thanks to its formalization using the Event-B language. We rely on this formalization approach to evaluate our taxonomy. The formalization activity helped us to correct and update our taxonomy as it evolves, as shown in part c of Figure~\ref{fig:process}. For instance, in the very beginning, we modeled authenticity as one of the first-level security properties according to the literature. When formalizing the integrity security property, we discovered that authenticity is in fact a refinement of integrity. Indeed, according to their definitions, the ``identity resource" in authenticity refines the ``data asset" in integrity, and ``spoofed" in authenticity refines ``misused" in integrity. 

Even though the use of taxonomy formalization helped evaluate its correctness, our approach presents the following limitations: 
\begin{itemize}
    \item \textit{Taxonomy updates.} Our work is not fully automated. This means that when we modify the hierarchy of some properties, we need to come back for updating manually the formalized corresponding Event-B models. We are aware of this difficulty. As a future direction, we are working on the use of Continuous Integration (CI) in an MBSE approach to take into account recurrent changes in our taxonomy. CI techniques promote the automated integration of changes. In fact,  this is a DevOps good practice that enables multiple developers to contribute and collaborate in a shared repository such as GitHub. As explained in this work, we needed to update our taxonomy to better reorganize the hierarchy of security properties. In some cases, this required a laborious manual effort for coordinating the update of taxonomy and the update of properties' formalization in parallel. The adoption of tools like Jenkins\footnote{\url{https://www.jenkins.io/}} or Bitbucket Pipelines\footnote{\url{https://bitbucket.org/product/features/pipelines}} will be  helpful to automate the updates. This work is one of the perspectives of this paper. 
    \item \textit{Non-exhaustive Validation.} The validation of our taxonomy models is based on scenarios extracted from the CAPEC database. Due to space limitations, we do not show all the animated scenarios. In addition, for some properties, like Authenticity, we faced some difficulties to find scenarios to check our models. In the CAPEC repository, an indication of scenarios consists of ``Employ robust authentication processes (e.g., multi-factor authentication)". As we can see, we need to improve the set of scenarios using other repositories. Besides, we plan to solicit security experts from the industry and propose brainstorming exercises for cybersecurity school students to enrich scenarios and to help validate the taxonomy models.
\end{itemize}

Besides, even if this taxonomy includes many types of factors that can impact a system's security, such as software, hardware, supply chain, and human aspects, the human confidence property is yet not much developed in this taxonomy. This is due to the security repositories that we used to extract security properties. Indeed, CAPEC and ATT\&CK contain rich information about technical attacks, but not social engineering attacks. To extend our taxonomy by adding social aspects, we intend to study existing works~\cite{wu2022sok, beckers2016serious, jeong2019towards} for a better understanding of human factors in cybersecurity, which is also our future work.

%% file: 8conclusion.tex
\section{Conclusion}
\label{sec:conclusion}
In this paper, we present a taxonomy of security properties, together with the methodology that we have used for defining it. This taxonomy provides an SDLC support for aligning security properties refinement with attack and defense refinement. It addresses ``what'' security properties a system or a system element should preserve rather than ``how" to implement them. 
To verify the proposed taxonomy, we have used the Event-B formal language to early check its correctness. We first formalized each property into Contexts and Machines and then verified and validated them using available tools included in the Rodin platform and validation scenarios extracted from CAPEC and ATT\&CK. To the best of our knowledge, this is the first SDLC taxonomy of formalized security properties provided in the literature. As a perspective of this work, we will bridge the proposed taxonomy with lower-level implementation. We are also working on enriching the database of the animation scenarios by looking further into other databases such as CWE. Moreover, sub-properties of human confidence will be included to extend this taxonomy, by leveraging social engineering attack knowledge.